\documentclass[twocolumn,prb,floatfix,amsmath,amssymb]{revtex4-1}
\setcitestyle{numbers,square}
\usepackage{amsfonts}%
\usepackage{mathrsfs,bbm}%
\usepackage{mathtools}%
\usepackage{xcolor}
\usepackage{graphicx}%
  \DeclareGraphicsExtensions{.pdf,.eps,.png}
\usepackage{prettyref}%
	\newcommand{\pref}[1]{\prettyref{#1}}%
	\newrefformat{fig}{Fig.~\ref{#1}}%
	\newrefformat{tab}{Table~\ref{#1}}%
	\newrefformat{sec}{Sec.~\ref{#1}}%
	\newrefformat{eq}{Eq.~(\ref{#1})}%

\newcommand{\qe}{\textsc{Quantum~ESPRESSO}}
\newcommand{\adeg}[1]{\ensuremath{#1^{\circ}}}
\newcommand{\im}{{\textrm{i}}}
\newcommand{\imt}[1]{\operatorname{Im} \left[ #1 \right]}
\newcommand{\bfR}{\ensuremath{\textbf{R}}}%
\newcommand{\ang}{\ensuremath{\textnormal{\,\AA}}}
\newcommand{\ev}{\ensuremath{{\textrm{\,eV}}}}%
\newcommand{\mev}{\ensuremath{{\textrm{\,meV}}}}%
\newcommand{\dxz}{\ensuremath{d_{xz}}}%
\newcommand{\dyz}{\ensuremath{d_{yz}}}%
\newcommand{\dxy}{\ensuremath{d_{xy}}}%
\newcommand{\sto}{SrTiO$_3$}%
\newcommand{\svo}{SrVO$_3$}%
\newcommand{\cvo}{CaVO$_3$}%
\newcommand{\ttg}{\ensuremath{t_{2g}}}
\newcommand{\atet}{\ensuremath{a_{\mathrm{tetr}}}}
\newcommand{\umit}{\ensuremath{U_{\textrm{\sc mit}}}}

\begin{document}

\title{Metal-insulator transition in CaVO$_3$ thin films: interplay
  between epitaxial strain, dimensional confinement, and surface
  effects} \date{\today}
\author{Sophie Beck}
\thanks{Both authors have contributed equally to this work.}
\author{Gabriele Sclauzero}
\thanks{Both authors have contributed equally to this work.}
\author{Uday Chopra}
\author{Claude Ederer}
\email{claude.ederer@mat.ethz.ch}
\affiliation{%
Materials Theory, ETH Zurich, Wolfgang-Pauli-Strasse 27, CH-8093 Z\"u{}rich, Switzerland}
\begin{abstract}
We use density functional theory plus dynamical mean-field theory
(DFT+DMFT) to study multiple control parameters for tuning the
metal-insulator transition (MIT) in \cvo\ thin films. We focus on
separating the effects resulting from substrate-induced epitaxial
strain from those related to the reduced thickness of the film. We
show that tensile epitaxial strain of around 3-4\% is sufficient to
induce a transition to a paramagnetic Mott-insulating phase. This
corresponds to the level of strain that could be achieved on a
\sto\ substrate. Using free-standing slab models, we then demonstrate
that reduced film thickness can also cause a MIT in \cvo, however,
only for thicknesses of less than 4 perovskite units. Our calculations
indicate that the MIT in such ultra-thin films results mainly from a
surface-induced crystal-field splitting between the \ttg-orbitals,
favoring the formation of an orbitally-polarized Mott insulator. This
surface-induced crystal-field splitting is of the same type as the one
resulting from tensile epitaxial strain, and thus the two effects can
also cooperate. Furthermore, our calculations confirm an enhancement
of correlation effects at the film surface, resulting in a reduced
quasiparticle spectral weight in the outermost layer, whereas
bulk-like properties are recovered within only a few layers away from
the surface.
\end{abstract}

\maketitle

\section{Introduction}
\label{sec:intro}

Transition-metal oxides offer an interesting playground, of both
technological and fundamental importance, which allows to engineer
materials properties through the fine-tuning of both structure and
stoichiometry~\cite{Cheong:2007,Dagotto/Tokura:2008}. Electronic,
magnetic, and structural properties of these materials are often
strongly coupled, due to the localized nature of the $3d$ electrons of
the transition metal cation, which form narrow bands around the Fermi
level dominating the electronic properties. This can lead to a number
of exotic phenomena such as metal-insulator transitions (MITs),
colossal magneto-resistance, or high-temperature superconductivity
\cite{Imada/Fujimori/Tokura:1998}.

Progress in modern thin film growth techniques has opened up
additional possibilities for tuning materials properties by growing
oxide heterostructures, such as superlattices or thin films, with
essentially atomically sharp interfaces~\cite{Mannhart/Schlom:2010}.
The physical properties of such heterostructures can be tuned by
choosing the appropriate combination of constituents, as well as the
super-periodicity or the film thickness.
Furthermore, within the interfacial region of these heterostructures,
completely novel properties can emerge, that are usually not found in
the corresponding bulk constituents~\cite{Hwang_et_al:2012}. A
striking example for such an emerging interfacial property is the
high-mobility two-dimensional electron gas observed at the interface
of the two insulators LaAlO$_3$ and
SrTiO$_3$~\cite{Ohtomo/Hwang:2004}.

In fact, such dramatic changes of the conductive properties compared
to the corresponding bulk materials are not necessarily restricted to
the interfacial region. Examples are the MITs observed in thin films
of various transition metal oxides, including, e.g.,
vanadates~\cite{Yoshimatsu_et_al:2010,Gu/Wolf/Lu:2014,Gu_et_al:2013},
titanates~\cite{Wong_et_al:2010,He_et_al:2012},
manganates~\cite{Liao_et_al:2015},
nickelates~\cite{Liu_et_al:2010,Scherwitzl_et_al:2011,Catalano_et_al:2014,Wang_et_al:2015},
or iridates~\cite{Biswas/Kim/Jeong:2014}.
Several of these materials are metallic in their bulk forms (e.g.,
SrVO$_3$, CaVO$_3$, La$_{2/3}$Ca$_{1/3}$MnO$_3$, LaNiO$_3$, or
SrIrO$_3$) and become insulating when the film thickness is reduced
below a certain critical value. The source for this MIT is often
ascribed to confinement effects, which restrict electron movement to
the in-plane directions of the film and lead to the formation of
quantum well
states~\cite{Okamoto:2011,Zhong/Zhang/Held:2013}. However, the
reported critical thicknesses for different materials are rather
wide-spread, e.g., 3 respectively 5 unit cells for
La$_{2/3}$Ca$_{1/3}$MnO$_3$~\cite{Liao_et_al:2015} and LaNiO$_3$
\cite{Scherwitzl_et_al:2011}, compared to 4\,nm ($\approx 11$ unit
cells) in CaVO$_3$~\cite{Gu_et_al:2013}. Furthermore, even for
different samples of the same material, the reported critical
thicknesses can differ quite significantly, e.g., 2-3 unit cells for
SrVO$_3$ in Ref.~\onlinecite{Yoshimatsu_et_al:2010}, compared to
6.5\,nm ($\approx 17$ unit cells) in
Ref.~\onlinecite{Gu/Wolf/Lu:2014}. This indicates that factors other
than dimensional confinement need to be taken into account to fully
understand the origin of such thickness-dependent MITs.

For example, it has been suggested, based on first-principles
electronic structure calculations, that early transition metal
perovskites such as SrVO$_3$ and CaVO$_3$ can become insulating under
strong tensile epitaxial
strain~\cite{Sclauzero/Dymkowski/Ederer:2016}. Such epitaxial strain
is often present in thin film samples due to the lattice mismatch
between the thin film and substrate materials. It has also been
reported that the critical thickness for the MIT in
La$_{2/3}$Sr$_{1/3}$MnO$_3$ is increased by the presence of both
strain and oxygen vacancies~\cite{Liao_et_al:2015}, and that a
transition from metallic to insulating behavior can be induced in
SrIrO$_3$ both by reducing the film thickness or by applying
compressive strain through a suitably chosen
substrate~\cite{Biswas/Kim/Jeong:2014}. The opposite effect, i.e., a
transition from a bulk Mott insulator to a metallic thin film, has
been reported for LaTiO$_3$ \cite{Wong_et_al:2010,He_et_al:2012}, and
attributed to the effect of compressive epitaxial
strain~\cite{Dymkowski/Ederer:2014}. Similarly, it has been found that
the transition to the low temperature insulating state in bulk
NdNiO$_3$ and SmNiO$_3$ can be completely suppressed by epitaxial
strain~\cite{Liu_et_al:2010,Catalano_et_al:2014}, and a competition of
strain (in this case favoring the metallic state) and dimensionality
effects (favoring the insulating state) has been reported for
NdNiO$_3$~\cite{Wang_et_al:2015}.

Here, we focus on the case of \cvo, which in its bulk form exhibits
relatively high resistivities \cite{Falcon_et_al:2004}. Furthermore,
its spectral properties point towards a strongly correlated metal
\cite{Nekrasov_et_al:2005,Yoshida_et_al:2010}, making it an ideal
candidate to probe the sensitivity of the Mott MIT to strain and
dimensionality effects.
Indeed, as already mentioned above, recent experiments have reported a
dramatic thickness-dependent change in the conducting properties of
\cvo\ films grown on \sto\ \cite{Gu_et_al:2013}. The sheet resistance
at low temperatures was shown to increase by orders of magnitude with
decreasing film thickness, and the temperature dependence of the
resistivity changed from metallic, for films thicker than 4\,nm, to
insulating, for the thinnest films.

In Ref.~\onlinecite{Gu_et_al:2013}, the thickness-dependent MIT has
been attributed to a reduction of the effective bandwidth, resulting
from a dimensional crossover from 3D to 2D. However, an alternative
scenario for the observed MIT, related to a thickness-dependent
relaxation of epitaxial strain, is outlined in the following.
In principle, large tensile epitaxial strain could be achieved in
\cvo\ films grown on a \sto\ substrate, with a nominal lattice
mismatch of 3.5\,\%. However, the out-of-plane lattice parameters
measured in an earlier study by Liberati {\it et
  al.}~\cite{Liberati_et_al:2009} indicate that for such a large
mismatch, the films tend to relax (at least partially) back toward the
bulk \cvo\ lattice parameters. Strain relaxation has also been
confirmed for a 53\,nm thick film investigated in
Ref.~\onlinecite{Gu_et_al:2013}. However, the strain states of the
thinner films have not been characterized. Typically, fully strained
epitaxial films can be grown up to a materials- and strain-dependent
critical thickness, above which the energy cost for forming misfit
dislocations becomes less than the elastic energy of a coherently
strained film \cite{Freund:1992}.
It is thus possible, that in very thin films, \cvo\ is essentially
fully strained to the in-plane lattice constant of the
\sto\ substrate. Such strong tensile strain can induce a transition to
insulating behavior, as predicted in
Ref.~\onlinecite{Sclauzero/Dymkowski/Ederer:2016}. For increasing film
thickness, the epitaxial strain is then successively released,
recovering the bulk-like metallic behavior in thicker films. We note
that for the rather large lattice mismatch between \cvo\ and \sto, the
critical thickness for the formation of misfit dislocations could be
rather small.

In this article, we assess the likelihood of this scenario, using
first principles-based electronic structure calculations. In
particular, we clarify the roles of both strain and reduced
dimensionality on the MIT in \cvo\ and related materials by first
incorporating each effect individually before treating them together
within our calculations.
For a realistic description of the MIT, we use a combination of
density functional theory and dynamical mean-field theory (DFT+DMFT)
\cite{Georges_et_al:1996,Anisimov_et_al:1997,Lechermann_et_al:2006,Held:2007},
which allows to obtain the Mott-insulating state without the need to
artifically introduce a symmetry-broken state with magnetic and/or
orbital long range order. In addition, the DFT+DMFT method accounts
well for correlation effects within the metallic state, as
demonstrated, e.g., for bulk \cvo\ and the related
\svo\ \cite{Pavarini_et_al:2004,Nekrasov_et_al:2005}.

The remainder of this article is organized as follows. After
describing our computational method in Sec.~\ref{sec:methods}, we
first confirm the prediction of
Ref.~\onlinecite{Sclauzero/Dymkowski/Ederer:2016}, and show that
tensile strain slightly above 3\% is sufficient to induce a MIT in
\cvo\ (Sec.~\ref{sec:strain}). We isolate the individual effect of
tensile epitaxial strain by using bulk unit cells with constrained
``in-plane'' cell parameters, thereby mimicking the elastic constraint
present in epitaxial films.
Then, in Sec.~\ref{sec:slabs}, we explore finite size effects within
strain-free slab geometries, i.e., corresponding to free-standing
ultra-thin films with thicknesses of 2-6 perovskite units. Indeed, we
find the thinnest slabs to be insulating. However, the metallicity is
recovered already for a thickness of 4 perovskite units, i.e.,
corresponding to film thicknesses of about 1.5\,nm. The
Mott-insulating character of the ultra-thin slabs can be traced back
to a surface-induced crystal-field splitting, similar to what has been
suggested recently from DFT+DMFT calculations for 2 unit cells of
\svo\ sandwiched between vacuum and a \sto\ ``substrate''
\cite{Zhong_et_al:2015}. The surface-induced crystal-field splitting
also leads to enhanced correlation effects in the surface layer of the
thicker \cvo\ slabs, indicated, e.g., by a large quasiparticle mass
enhancement. This is in good agreement with experimental photoemission
results \cite{Maiti_et_al:2001,Eguchi_et_al:2006} and previous
DMFT-based studies for \svo\ \cite{Liebsch:2003}.
Finally, we show that strain and finite-size effects cooperate, but
that the strain effect is clearly dominating, at least for thicknesses
above 4 perovskite units. This suggests that the scenario described
above, i.e., successive strain relaxation with increasing film
thickness as main mechanism behind the observed thickness-dependent
MIT in \cvo, is indeed highly plausible. Further experimental
characterization is required to verify this scenario.
Our work shows that it is crucial to take into account the interplay
of a variety of factors, such as, e.g., epitaxial strain, dimensional
confinement, defects, as well as genuine interface or surface effects,
in order to fully understand the emerging properties observed in
complex oxide heterostructures.

\section{Computational Method}\label{sec:methods}

We use the \qe\ package \cite{Giannozzi_et_al:2009} together with the
PBE functional \cite{Perdew/Burke/Ernzerhof:1996} to calculate the
electronic structure within DFT and perform all structural
relaxations.
To address the effect of epitaxial strain, we use the bulk $Pbnm$ unit
cell of CaVO$_3$ and constrain the orthorhombic ``in-plane'' lattice
constants $a$ and $b$ to be equal to a given $a_\text{substrate}$,
while relaxing the ``out-of-plane'' lattice parameter $c$ and all
internal structural degrees of freedom (atomic positions). Thus, we
use a geometry where the long orthorhombic axis is perpendicular to
the plane defined by the substrate surface, which preserves the
orthorhombic $Pbnm$ symmetry ({\it c.f.}
Ref.~\onlinecite{Sclauzero/Ederer:2015}). The epitaxial strain is then
defined as $s = (a_\text{substrate} - a_0)/a_0$, where $a_0$ is the
in-plane lattice constant corresponding to the unstrained system,
which we choose as the average of $a$ and $b$ in the fully relaxed
bulk system.

\begin{figure}
  \includegraphics[width=\columnwidth,height=0.3\textheight,keepaspectratio]{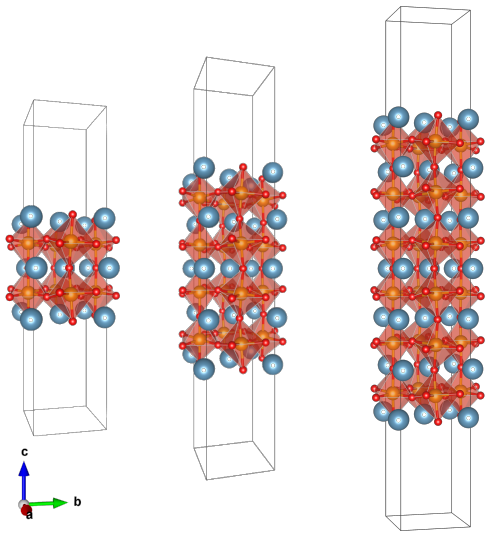}
  \caption{(Color online) Unit cells for the 1-, 2- and 3-bilayer slab
    calculations (from left to right). Ca atoms are shown in blue, V
    in orange, and O in red. Each slab consists of $2l$ perovskite
    units, terminated on both sides by CaO layers, and separated by
    16\,\ang\ of vacuum.}\label{fig:SLABS}
\end{figure}

To address the finite size effect, we use slab supercells consisting
of 2$l$ perovskite units, $l \in \{1,2,3\}$, stacked along the $z$
direction and terminated by CaO layers on both sides, which are
separated by a sufficient amount of vacuum, as shown in
\pref{fig:SLABS}.
The slabs are built starting from the $Pbnm$ bulk structure and thus
employ $\sqrt{2} \times \sqrt{2}$ lattice vectors (relative to the
simple perovskite unit) perpendicular to $z$, such as to allow for the
same octahedral tilt pattern as in the $Pbnm$ bulk structure.
This slab geometry preserves the mirror plane $m$ perpendicular to $z$
within the central CaO layer of the slab, as well as the glide plane
$b$ parallel to $z$, while the glide plane $n$ is not preserved (since
$b$ has its glide vector perpendicular to $z$ which is not the case
for $n$). Thus, the upper and lower halfs of the slabs are
symmetry-equivalent due to $m$, and the two V sites within each layer
are symmetry-equivalent due to $b$, resulting in only $l$ inequivalent
V sites per slab.
We use the same lattice constant $a_0$ along both in-plane directions
of the slab, analogous to the unstrained bulk reference state. In
order to assess the combined effect of strain and finite size, we also
perform calculations for ``epitaxially strained'' slabs, where we
constrain the in-plane lattice constant to values larger than $a_0$.
During the structural relaxation of the slabs, the atomic positions
within the two inner layers are fixed to the corresponding bulk
structures (at the given strain), while all other positions (present
only for $l>1$) are relaxed. Even for $l=1$, the outermost CaO layer
is also relaxed. We found that a vacuum layer of 16\,\ang\ is necessary
and sufficient to avoid cross-talk between the periodic images along
$z$.

We use standard PBE scalar-relativistic ultrasoft pseudopotentials
from the \qe\ website, where the $3s$ and $3p$ semicore
states of both Ca and V are treated as valence electrons. The
plane-wave energy cutoffs for the expansion of wave-functions and
charge density are set to 40 and 300 \,Ry, respectively. Brillouin
zone integrals are evaluated using the Methfessel-Paxton method with a
smearing parameter of 0.02\,Ry and a $6\times 6 \times 4$ $k$-point
grid for the $Pbnm$ unit cell of CaVO$_3$, which is reduced to a
$6\times 6\times 1$ grid for all slab unit cells.

After relaxing the structure within PBE-DFT, we construct maximally
localized Wannier functions (MLWFs) \cite{Marzari_et_al:2012} for the
\ttg-derived bands around the Fermi level, using the wannier90 code
\cite{Mostofi_et_al:2008}. We obtain three V-centered \ttg-like MLWFs
per V atom, which we use as basis orbitals to describe the low-energy
correlated subspace of \cvo\ ({\it c.f.}
Ref.~\onlinecite{Pavarini_et_al:2005}). The electron-electron
interaction within this correlated subspace is then treated within
DMFT to obtain local Green's functions.

The DMFT calculations are implemented using the TRIQS/DFTTools
libraries \cite{Parcollet_et_al:2015,Aichhorn_et_al:2016}. In the DMFT
calculations for the slab unit cells, a separate impurity problem is
solved for each of the inequivalent V sites (one for each layer $l$),
using the TRIQS/CTHYB solver
\cite{Gull_et_al:2011,Seth_et_al:2016}. The different impurity
problems are then connected through the self-consistency condition for
the lattice Green's function. We use the Slater-Kanamori form,
including spin-flip and pair-hopping terms, to describe the local
Coulomb interaction between the electrons on the impurity sites, and
we include all off-diagonal elements of the local Green's function and
self-energy in the calculation (after transforming to a coordinate
system where the local MLWF Hamiltonian is diagonal). All DMFT
calculations are performed for an inverse electronic temperature of
$\beta = 1/(k_\text{B}T) = 40\ev^{-1}$, which corresponds to
approximately room temperature.
From the imaginary-time Green's function, $G(\tau)$, we obtain the
orbital occupations, \mbox{$n = -G(\beta)$}, and the ``averaged''
spectral density at the Fermi level, \mbox{$\bar{A}(0) = -
  \tfrac{\beta}{\pi} \textrm{Tr}[G(\beta/2)]$}, which approaches
$A(\omega = 0)$ for decreasing temperature, i.e., in the limit $\beta
\rightarrow \infty$ ({\it c.f.} \cite{Fuchs_et_al:2011}).
The full real-frequency spectral functions, $A(\omega)$, are obtained
from $G(\tau)$ using the maximum entropy method
\cite{Jarrell/Gubernatis:1996}.

\section{Effect of epitaxial strain}
\label{sec:strain}

\begin{table}[t]
  \caption{Orthorhombic lattice parameters $a$, $b$, and $c$ (in \ang)
    and octahedral rotation and tilt angles $\theta$ and $\phi$ (in
    degrees) for the $Pbnm$ paramagnetic bulk phase obtained within
    this work (DFT) and from neutron diffraction experiments at room
    temperature (Exp.)~\cite{Falcon_et_al:2004}. The averaged
    pseudo-tetragonal cell parameter defined as
    $\atet=(a+b)/2\sqrt{2}$ is used as reference value to define zero
    epitaxial strain.}  \setlength{\tabcolsep}{5pt}
  \begin{tabular}{ccccccc}
    \hline\hline
      &   $a$  &   $b$  &   $c$  &  \atet & $\theta$  &  $\phi$  \\ \hline
 DFT  &  5.319 &  5.365 &  7.554 &  3.777 &   7.9     &   9.4   \\ 
 Exp. &  5.322 &  5.343 &  7.547 &  3.771 &   7.9     &   9.3    \\ \hline\hline
  \end{tabular}
  \label{tab:cell}
\end{table}

\begin{figure}
  \includegraphics{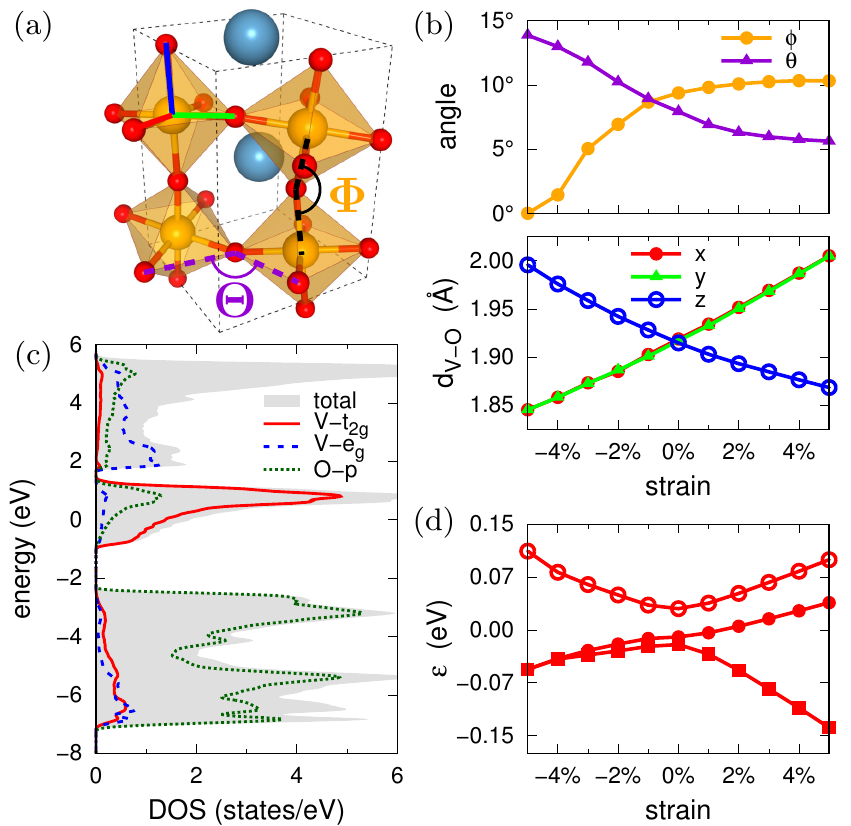}
  \caption{(Color online) DFT results for strained bulk \cvo. (a)
    V--O--V and O--O--O bond angles $\Phi$ and $\Theta$ used to define
    the octahedral tilt and rotation angles $\phi=(\pi-\Phi)/2$ and
    $\theta = (\pi/2-\Theta)/2$ shown in the upper panel of (b) as
    function of epitaxial strain. The lower panel of (b) shows the
    strain dependence of the V--O bond distances, labeled according to
    the approximate orientation of the corresponding bond (also
    indicated with the same colors in (a)). (c) Total and
    V-$t_{2g}$/$e_g$/O-$p$-projected densities of states (DOS) for
    zero strain with the Fermi level at zero energy. (d) Strain
    dependent crystal-field levels $\varepsilon$ obtained from the
    eigenvalues of the $3 \times 3$ on-site Hamiltonian matrix in the
    MLWF basis (here, $\varepsilon=0$ is taken as the average level
    energy at each strain value).}.
  \label{fig:bulkDFT}
\end{figure}

We first relax the bulk $Pbnm$ structure (cell parameters as well as
atomic positions) without applying any constraints within DFT. To
characterize the V--O--V bond angles in the relaxed structures, we use
the conventional rotation and tilt angles, $\theta$ and $\phi$,
corresponding to the $a^- a^- c^+$ tilt system (see
\pref{fig:bulkDFT}a and, e.g., Ref.~\cite{Rondinelli/Spaldin:2011}).
As seen from \pref{tab:cell}, these angles as well as the relaxed
lattice parameters are in excellent agreement with experimental data
obtained from neutron diffraction \citep{Falcon_et_al:2004}.

We then study the effect of epitaxial strain on the crystal structure
of \cvo\ for the case of a square-lattice substrate (e.g., \sto) by
constraining the in-plane cell parameters while relaxing all other
structural degrees of freedom, as explained in \pref{sec:methods}.
The theoretical bulk structure has a ratio of $b/a=1.009$, slightly
different from 1, so we choose to refer the nominal strain value to
the average in-plane spacing
\mbox{$\atet=(a+b)/2\sqrt{2}=a_0/\sqrt{2}=3.777\ang$}.

The application of strain strongly influences the internal degrees of
freedom of the crystal structure, especially bond angles and bond
distances. The variation of the two conventional rotation and tilt
angles, as well as of the V--O bond lengths, are shown in
\pref{fig:bulkDFT}b as function of strain. Under compressive strain,
the out-of-plane bonds show a strong tendency to straighten, leading
to a complete suppression of tilts for compressive strains stronger
than $-$4\,\%.  Simultaneously, the rotation angle $\theta$ increases
significantly with respect to the unstrained case. Under tensile
strain, the opposite trend is observed, i.e., increasing out-of-plane
tilts and decreasing in-plane rotations. However both angles tend to
saturate towards large tensile strain, to values of around \adeg{10.3}
and \adeg{5.6}, respectively, i.e., only slightly larger, respectively
smaller, than the bulk values.
Regarding the V--O bond lengths, we note that the in-plane V--O bond
lengths follow the applied strain, while the out-of-plane bond length
exhibits the opposite trend, dictated by the expansion (contraction)
of the out-of plane cell parameters in response to compressive
(tensile) strain. These trends are similar to what has been reported
for other early transition metal perovskite oxides (see, e.g.,
Refs.~\cite{Dymkowski/Ederer:2014,Sclauzero/Ederer:2015}).

The electronic density of states (both total and projected on atomic
V-\ttg/$e_g$ and O-$p$ orbitals) for unstrained \cvo\ is shown in
\pref{fig:bulkDFT}c. It can be seen, that a metallic state is obtained
within DFT, and that the electronic states around the Fermi level are
formed by a group of bands with dominant V-\ttg\ character, which is
separated from other bands at higher and lower energies. The densities
of states for the strained structures are qualitatively similar to
that.
For each of these structures, we now construct MLWFs corresponding to
the V-\ttg-dominated bands around the Fermi energy. As for other early
transition metal oxides (see, e.g.,
Refs.~\cite{Lechermann_et_al:2006,Dymkowski/Ederer:2014}), we obtain
three MLWFs per V site, which resemble atomic \ttg\ orbitals, albeit
with strong $p$-like tails on the surrounding oxygen ligands.

We then calculate the matrix elements of the Kohn-Sham Hamiltonian,
$h_{ij}[\bfR'] = \langle w_i[\bfR+\bfR'] | \hat{H} | w_j[\bfR]
\rangle$, where $i,j \in \{1,2,3\}$ are the orbital indeces of the
MLWFs and \bfR\ and $\bfR+\bfR'$ indicate different V
sites~\footnote{We note that here we are using a simplified notation
  where \bfR\ and \bfR' do not necessarily correspond to lattice
  translations of the $Pbnm$ structure. Thus, strictly speaking, these
  matrix elements can depend on both \bfR\ and $\bfR'$. However, since
  all V sites are equivalent, one can always find a local
  transformation such that the matrix elements for different
  \bfR\ become identical.}.
Thereby, orbital ``3'' generally resembles \dxy, i.e., it is almost
in-plane, while orbitals ``1'' and ``2'' resemble linear combinations
of \dxz\ and \dyz.
The crystal-field levels $\varepsilon_i$, obtained as eigenvalues of
the on-site $3 \times 3$ matrix $h_{ij}[\bfR' = (0,0,0)]$, are shown
as function of strain in \pref{fig:bulkDFT}d. They exhibit the same
trends that have already been observed in similar systems
\cite{Sclauzero/Dymkowski/Ederer:2016}. The epitaxial stain induces a
tetragonal-like splitting between the three \ttg-MLWFs that depends
approximately linear on the strain, with one orbital at higher and two
orbitals at lower energies under strong compressive strain, and vice
versa under strong tensile strain. Superimposed on this
tetragonal-like splitting is the effect of the octahedral tilts and
rotations, which lifts the degeneracy between the three orbitals
already for zero strain, and also splits the two higher lying levels
in the tensile strain regime. Under compressive strain, this additonal
splitting is small and vanishes completely below $-4$\,\% strain, due
to the suppression of the out-of-plane tilts which raises the symmetry
of the system to tetragonal.
Further, we note that the nearest neighbor hopping amplitudes exhibit
a similar strain dependence to what has been reported previously for
related systems \cite{Sclauzero/Dymkowski/Ederer:2016}.

\begin{figure}
  \includegraphics[width=1.0\columnwidth]{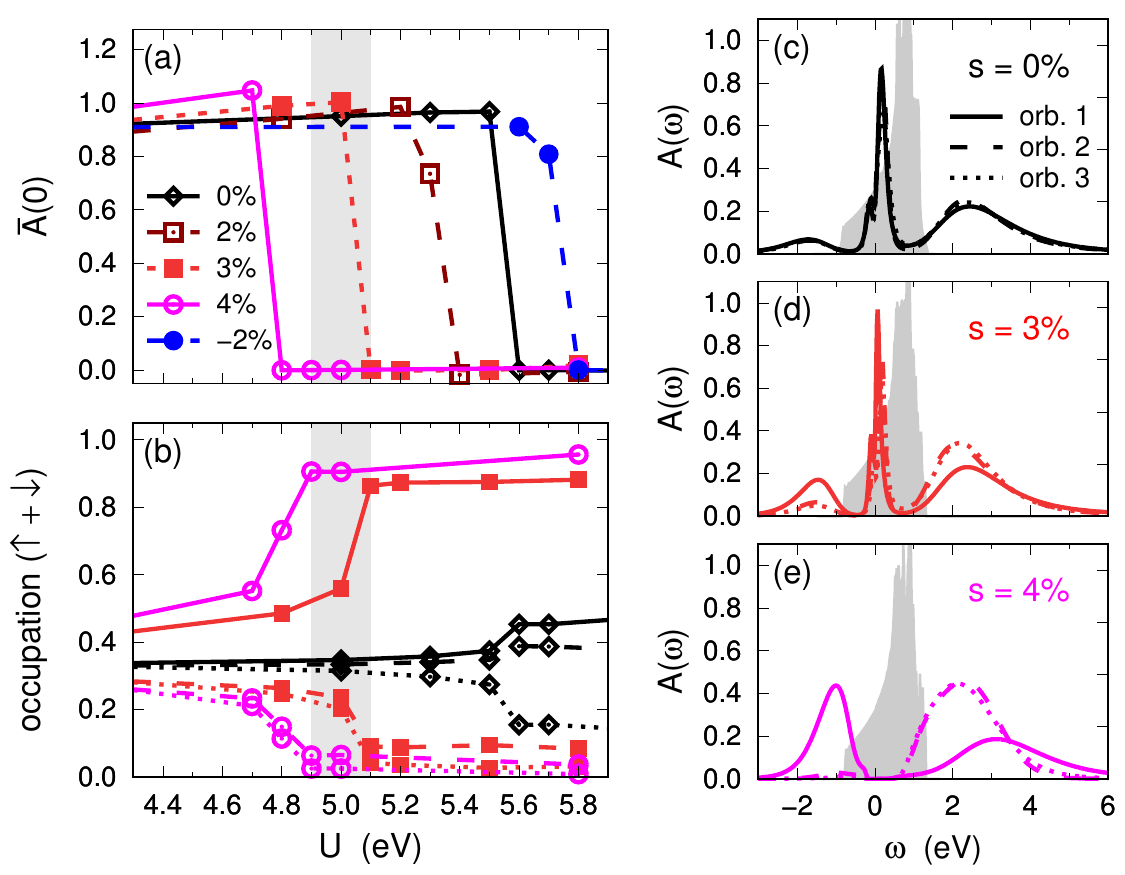}
  \caption{(Color online) DFT+DMFT results for strained bulk \cvo.
    (a) Averaged spectral density at the Fermi level $\bar{A}(0)$ and
    (b) orbital-resolved occupations from the eigenvalues of
    $-G(\beta)$, summed over both spin components, as function of the
    interaction parameter $U$ for different intensities of the
    epitaxial strain (symbols). In (b), only strains of 0\,\%, 3\,\%,
    and 4\,\% are shown.  (c-e) Orbital-resolved spectral function
    $A(\omega)$ for different values of strain calculated for
    $U=5\ev$.  In (b-e), different line styles (solid, dashed, dotted)
    indicate different orbitals.}\label{fig:bulkDMFT}
\end{figure}

Next, we investigate whether epitaxial strain facilitates the
formation of a Mott-insulating state in \cvo, by performing DMFT
calculations within the correlated subspace defined in the MLWF basis,
as described in \pref{sec:methods}. \pref{fig:bulkDMFT}a shows
$\bar{A}(0)$, the averaged spectral weight around the Fermi level, as
function of the interaction parameter $U$ for different values of
strain. For the unstrained case (black line) one observes a MIT at
$U\simeq5.5\ev$, i.e., the system is metallic ($\bar{A}(0) > 0$) for
$U<5.5\ev$ and insulating ($\bar{A}(0) \approx 0$) for larger $U$
values. Thus, for a typical value of $U=5\ev$, which has been used
previously for \cvo\ and related materials \cite{Pavarini_et_al:2004},
the system is metallic but not too far from the MIT, confirming the
nature of unstrained \cvo\ as ``correlated metal''.

Considering now the results for the strained systems, one observes
that epitaxial strain has a remarkable effect on the electronic
properties of \cvo.  Under tensile strain, the critical value of $U$
at which the MIT occurs is strongly shifted to lower values, reaching
the region around $U \approx 5 \ev$ for strains between 3\,\% and
4\,\%. In contrast, the critical $U$ for the MIT is increased under
compressive strain. These trends are in line with the general
considerations discussed in
Ref.~\cite{Sclauzero/Dymkowski/Ederer:2016}.
\mbox{\pref{fig:bulkDMFT}c-e} show the spectral functions calculated
for $U=5\ev$. One can see that a sizable gap of $\sim 1\ev$ opens in
the \ttg\ bands for a strain of about 4\,\% (\pref{fig:bulkDMFT}e),
while already at 3\,\% strain (\pref{fig:bulkDMFT}d) there is a
narrowing of the quasiparticle peak compared to the unstrained
case. These spectral changes are accompanied by noticeable changes in
the orbital occupations, as shown in \pref{fig:bulkDMFT}b.
In the unstrained case, all three \ttg-orbitals are equally occupied
in the metallic regime, and only a weak orbital polarization develops
in the insulating state, i.e. for high values of $U$. In contrast,
under 3\,\% and 4\,\% tensile strain, there is a sizeable orbital
polarization already in the metallic state, and the system becomes
essentially fully polarized, with one half-filled and two empty
orbitals, in the insulating regime. This is consistent with the
strain-induced crystal-field splitting shown in \pref{fig:bulkDFT}d,
and is analogous to what has been reported for other $d^1$ perovskites
in Refs.~\onlinecite{Sclauzero/Dymkowski/Ederer:2016} and
\onlinecite{Dymkowski/Ederer:2014}.

It follows that the level of strain achievable in \cvo\ thin films
using \sto\ as substrate (about 3.5\% tensile strain) is expected to
bring the system right at the boundary of a MIT (or even across it).

\section{Effect of dimensional confinement}
\label{sec:slabs}

Next, we investigate how confinement perpendicular to the substrate
plane, due to finite film thickness, affects the tendency of \cvo\ to
undergo a MIT. For this, we consider unstrained free-standing slab
unit cells of \cvo, with thicknesses varying from 2 to 6 simple
perovskite unit cells (corresponding to about 0.8\,nm to 2.2\,nm), as
described in \pref{sec:methods} and also shown in \pref{fig:SLABS}.
We note that, while here we are using the free-standing slab geometry
mainly as model systems to isolate finite size effects from other
factors affecting the MIT, a procedure to experimentally fabricate
such free-standing perovskite films has in fact been recently
reported~\cite{Lu_et_al:2016}.

\begin{figure}
  \includegraphics[width=1.0\columnwidth]{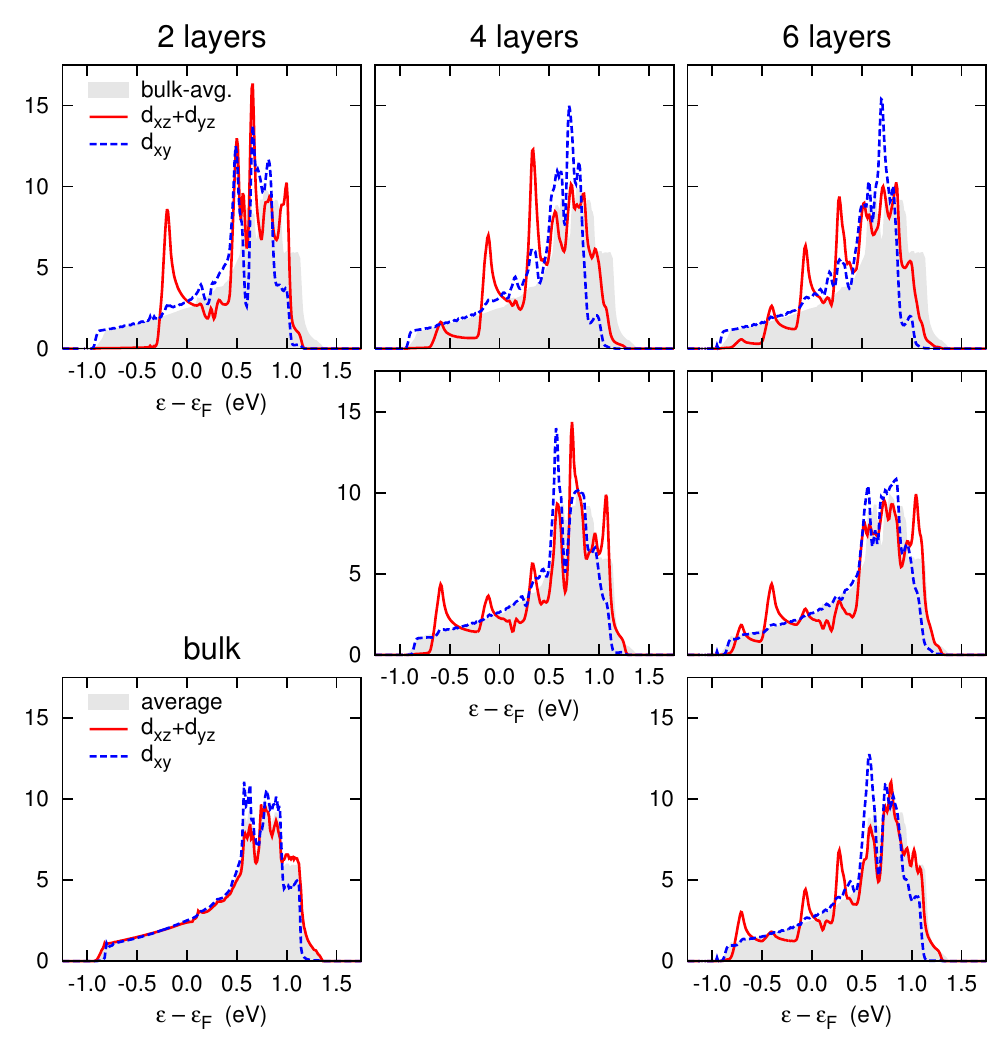}
  \caption{(Color online) Layer- and orbitally-resolved DOS
    corresponding to the \ttg-bands represented in the MLWF basis for
    unstrained slabs of different thicknesses (left to right). Each
    row corresponds to a different layer, from the surface (top)
    towards the innermost layer (bottom). The bulk case is shown in
    the bottom left corner. The filled grey curve in all panels
    represents the average over the three \ttg-orbitals in the bulk,
    while the dashed blue (solid red) lines correspond to the
    \dxy-like (average of \dxz- and \dyz-like) MLWFs.}
  \label{fig:pdos}
\end{figure}

After structural relaxation of the different slabs, as described in
\pref{sec:methods}, we construct MLWFs for the V-\ttg-dominated bands
around the Fermi level. Similar to the corresponding bulk system
(shown in \pref{fig:bulkDFT}c), these bands are well separated from
other bands at higher and lower energies, and the resulting MLWFs are
centered on the V sites with strong local \ttg\ character plus
additional O-$p$ contributions on the surrounding ligands.
Fig.~\ref{fig:pdos} shows the layer- and orbitally-resolved (in the
MLWF basis) densities of states (DOS) of these ``\ttg-bands'' for the
three different slabs with different thicknesses. For comparison, also
the corresponding DOS for the (unstrained) bulk system is shown.

One can see that, in the dimensionally-confined systems, there is a
clear difference between the DOS corresponding to the in-plane
\dxy-like MLWFs and the one corresponding to the other two orbitals,
whereas in the bulk case the DOS of all three MLWFs are very similar.
Due to its orientation within the $x$-$y$ plane and the nearly
two-dimensional character of the corresponding band, the \dxy-orbital
is not very sensitive to the confinement along $z$ and exhibits a
similar bandwidth as in the bulk. In contrast, the other two orbitals
have a significantly reduced bandwidth in the 1-bilayer slab, but also
(to a minor extent) in the 2- and 3-bilayer slabs. In all cases, the
corresponding DOS exhibits an evident subband structure, manifested as
multiple peaks, due to the confinement along the $z$ direction. The
quantum well states responsible for this subband structure are spread
throughout the slab, as can be seen from the matching peaks in the DOS
corresponding to different layers. This was described in more detail
for SrVO$_3$ slabs by \citet{Zhong/Zhang/Held:2013}.

\begin{figure}
  \includegraphics[width=1.0\columnwidth]{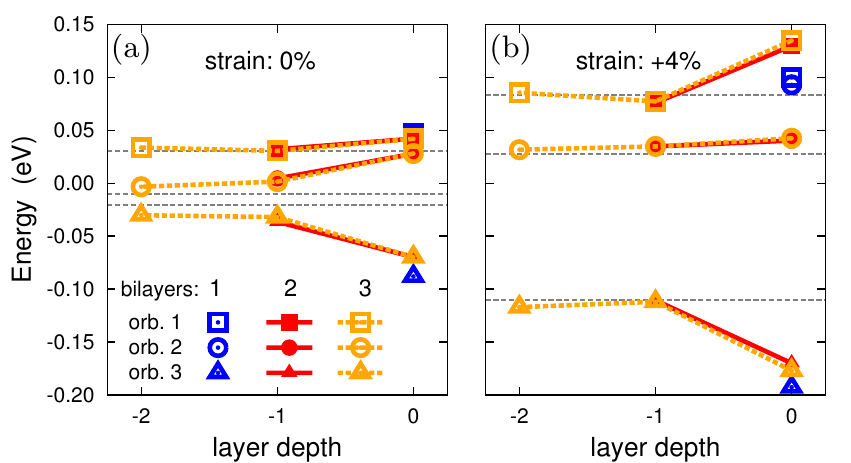}
  \caption{(Color online) Crystal-field splitting of the three
    \ttg-orbitals as a function of the layer depth in \cvo\ slabs with
    different number of bilayers (1, 2, and 3).  The left (right)
    panel shows the unstrained case (4\% tensile strain) and the
    corresponding bulk reference is indicated by the black dashed
    lines. Note that orbital 1 has dominant $d_{xy}$ character,
    whereas orbitals 2 and 3 are predominantly $d_{xz}$/$d_{yz}$. The
    average level energy is taken as zero in each layer.}
  \label{fig:CFS}
\end{figure}

We now examine the changes of the crystal-field levels for the
different slabs.  In \pref{fig:CFS}a, the crystal-field splitting
between the \ttg-orbitals is shown as a function of the layer depth
for the \mbox{1-,} 2-, and 3-bilayer slabs ($l=1,2,3$), together with
the bulk values as reference. It can be seen, that the results for
$l=2$ are nearly indistinguishable from the two outer layers of the
3-bilayer slab, and that the crystal-field levels in the inner layer
of the $l=3$ slab are already very similar to the bulk reference. This
shows that the influence of the surface decays very rapidly towards
the inside of the material and is nearly restricted to only the
outermost layer.
Furthermore, a strong crystal-field splitting is apparent in the
surface layer for all slab thicknesses. The \dxy-dominated orbital is
lowered in energy with respect to the other two orbitals, which in
turn become almost degenerate.
This is similar to what has been reported in
Ref.~\onlinecite{Zhong_et_al:2015} for ultra-thin SrVO$_3$, and
follows from the tetragonal-like symmetry-breaking at the surface.
The splitting between \dxy\ and \dxz/\dyz\ is largest in the 1-bilayer
(which consists only of two surface layers) and converges to about
110\,\mev\ already for $l=2$. The magnitude of this splitting is
comparable to the crystal-field splitting induced by a tensile strain
of about 3\% in the bulk system, and thus can be expected to have a
noticeable effect on the MIT.

We note that, in contrast to the local crystal-field energies, the
nearest neighbor hopping amplitudes (not shown) are only weakly
affected by the presence of the surface, and are overall very similar
to the bulk case (except, of course, that there is no hopping into the
vacuum layer).

\begin{figure}
  \includegraphics[width=0.8\columnwidth]{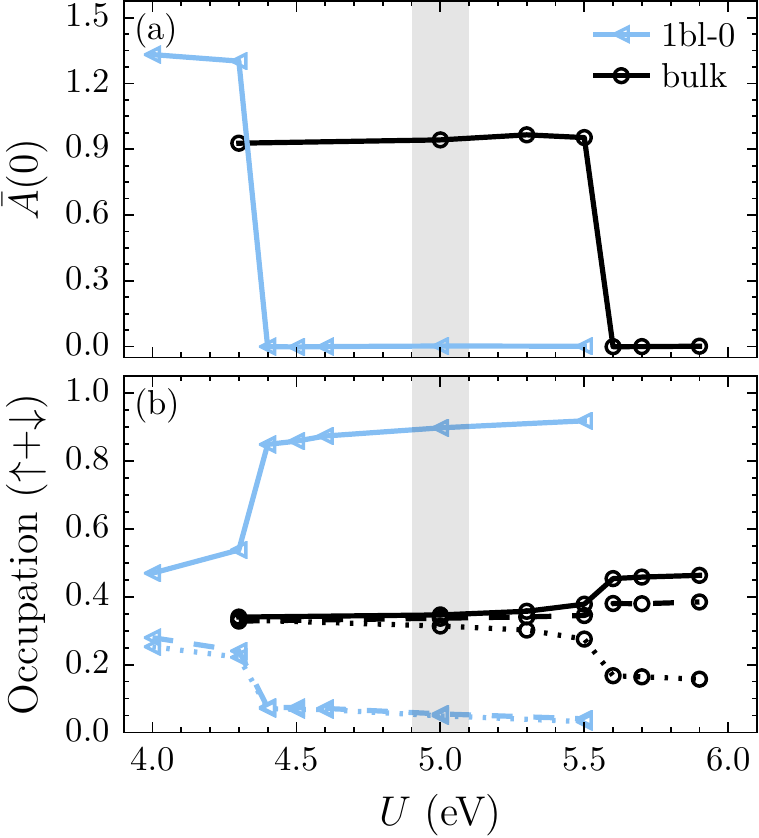}
   \caption{(Color online) DMFT results for the 1-bilayer \cvo\ slab
     in the unstrained case.  (a) Averaged spectral weight at
     $\omega=0$ and (b) orbitally-resolved occupations as a function
     of the interaction parameter $U$. Light blue lines and symbols
     refer to the slab, while black ones represent the bulk
     reference.}\label{fig:DMFT2l}
\end{figure}

We now perform DMFT calculations for the different slabs, using the
low-energy correlated subspace defined by the \ttg-MLWFs. We first
discuss the results for the 1-bilayer slab. From the evolution of
$\bar{A}(0)$, shown as function of the interaction strength $U$ in
\pref{fig:DMFT2l}a, it can be seen that there is a large shift of the
critical $U$ value for the MIT, \umit, from 5.6\ev\ in the unstrained
bulk case to 4.4\ev\ in the 1-bilayer slab.
Furthermore, from the corresponding orbital occupations shown in
\pref{fig:DMFT2l}b, one can see that the 1-bilayer slab is essentially
fully orbitally polarized in the Mott-insulating state, with one
half-filled and two empty orbitals, consistent with the strong
crystal-field splitting discussed above.
Considering a realistic $U$ value for the bulk system of around
5\ev\ (see \pref{sec:strain}), and taking into account that the $U$
value for the ultrathin slab is likely to be even somewhat larger due
to reduced screening, our results indicate that 1-bilayer of \cvo\ is
insulating, analogous to 1-bilayer of
\svo\ \cite{Yoshimatsu_et_al:2010,Zhong_et_al:2015}.

\begin{figure}
  \includegraphics[width=0.8\columnwidth]{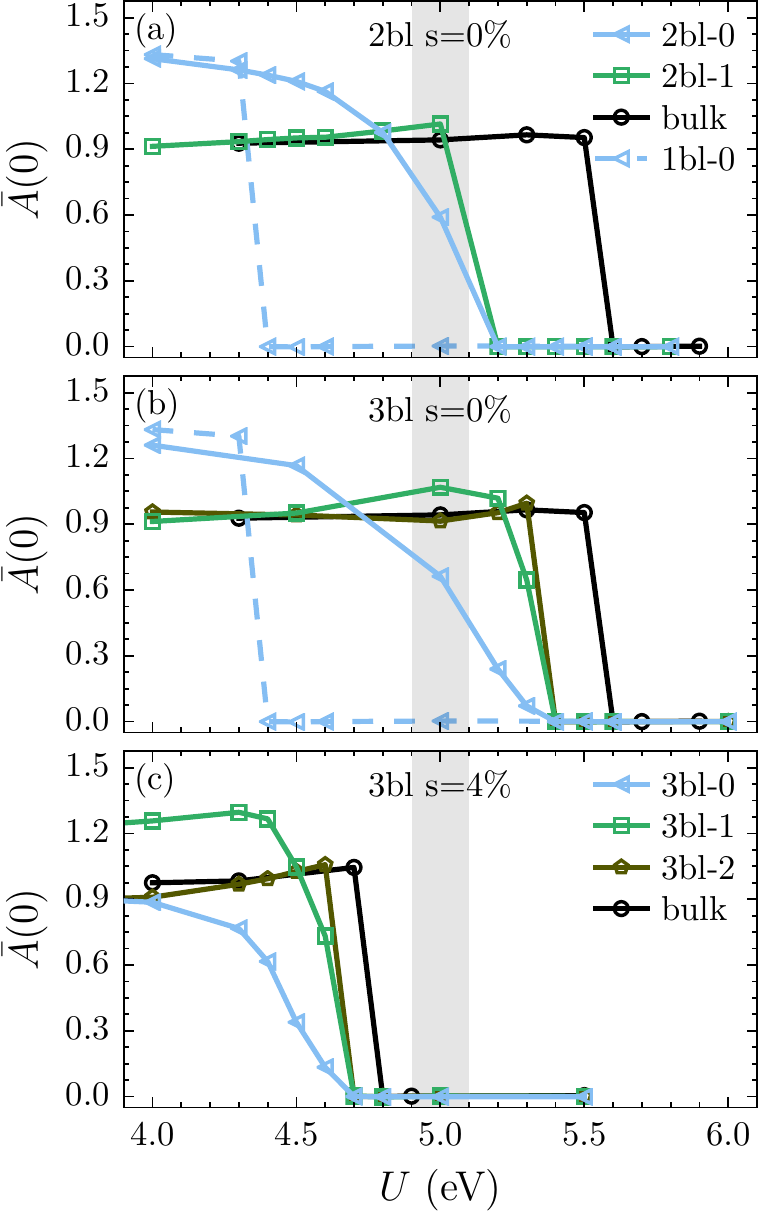}
  \caption{(Color online) Averaged spectral function at $\omega = 0$
    for the unstrained 2-bilayer (``2bl'') slab (a), the unstrained
    3-bilayer (``3bl'') slab (b), and the 3-bilayer slab at 4\% strain
    (c). In all panels the surface layer (``-0'') is indicated in
    light blue and using triangles, the subsurface layer (``-1'') in
    light green using squares, and the inner layer (``-2'', only
    present for 3bl) in dark green using pentagons. For comparison,
    the 1-bilayer (only in panels (a) and (b)) and the bulk reference
    (unstrained in (a) and (b), at 4\,\% tensile strain in (c)) are
    indicated by dashed blue and solid black lines,
    respectively.}\label{fig:DMFT46l}
\end{figure}

\begin{figure}
  \includegraphics[width=\columnwidth]{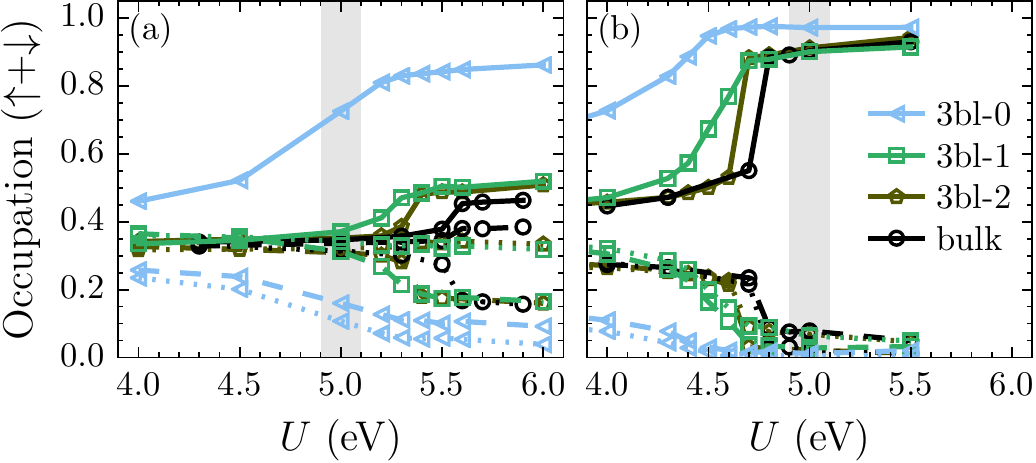} 
  \caption{(Color online) Orbital occupations as function of the
    interaction parameter $U$, calculated within DMFT for the
    3-bilayer slab in the unstrained case (a) and under 4\% tensile
    strain (b). Light to dark colors indicate the surface (3bl-0),
    subsurface (3bl-1), and inner (3bl-2) layers. The corresponding
    bulk data, unstrained (a) and at 4\,\% tensile strain (b), are
    shown in black.}\label{fig:OCC6l}
\end{figure}

Focusing now on the 2- and 3-bilayer slabs (\pref{fig:DMFT46l}a and
b), one can see that the shift of \umit\ quickly becomes weaker as the
slab thickness increases. For the 2-bilayer slab, shown in
\pref{fig:DMFT46l}a, \umit\ is only about 0.4\ev\ smaller than in the
bulk case, while for the 3-bilayer slab (\pref{fig:DMFT46l}b)
\umit\ is only 0.2\ev\ smaller than in the bulk.  Thus, our results
indicate that at $U=5\ev$, the 3-bilayer slab will already be
metallic, similar to the bulk system, while 2-bilayers are just on the
verge of the MIT.
However, in both cases, the surface layer exhibits a notably different
behavior compared to the inner or subsurface layers. Even though the
corresponding $\bar{A}(0)$ becomes zero at essentially the same $U$
value as the inner layers, the transition appears to be more
continuous in the surface layer. In addition, the occupations in the
surface layer exhibit strong orbital polarization already below
\umit\ (see \pref{fig:OCC6l}a for $U=5.0\ev$), consistent with the
large crystal-field splitting in the surface layer depicted in
\pref{fig:CFS}a, whereas the inner and subsurface layers are only
weakly polarized, similar to the bulk case.

\begin{figure}
  \includegraphics[width=\columnwidth]{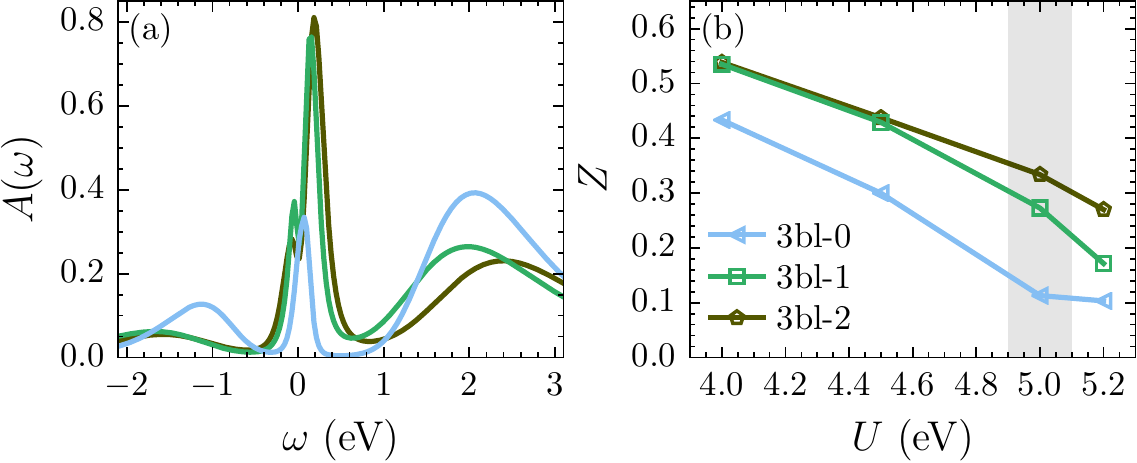} 
  \caption{(Color online) (a) Layer-resolved orbitally-averaged
    spectral functions $A(\omega)$ for the 3-bilayer slab and
    $U=5\ev$. (b) Inverse quasiparticle mass enhancement
    $\ensuremath{Z = \{m^*/m\}^{-1}}$ as function of the interaction
    parameter $U$. Light to dark colors in both (a) and (b) indicate
    the surface (3bl-0), subsurface (3bl-1), and inner (3bl-2) layers,
    as indicated by the legend.}\label{fig:SPEC6l}
\end{figure}

To further analyze the different behavior of the surface,
\pref{fig:SPEC6l}a shows layer-resolved spectral functions for the
3-bilayer slab at $U=5.0\ev$. It can be seen, that in particular the
inner, but also to a large extent the subsurface layer, exhibits a
spectral function that is very similar to the bulk (shown in
\pref{fig:bulkDMFT}c).
In contrast, the quasiparticle peak in the surface layer is much
narrower, and there is a stronger transfer of spectral weight into the
upper and lower Hubbard bands, which are significantly more pronounced
than for the subsurface and inner layers.
The strongly reduced spectral weight of the quasiparticle peak at the
surface also follows from the behavior of the inverse quasiparticle
mass enhancement, $Z$, shown as function of $U$ in \pref{fig:SPEC6l}b,
which is consistently smaller in the surface layer than in the other
two layers, and is nearly vanishing for $U=5\ev$. Here, $Z$ is
estimated from the imaginary part of the self-energy at the lowest
Matsubara frequency ({\it c.f.} \cite{Fuchs_et_al:2011}):
\begin{equation}
 Z = \left\{\frac{m*}{m} \right\}^{-1} \approx \left\{ 1 -
 \frac{\imt{\Sigma (\im \omega_0)}}{\omega_0} \right\}^{-1} \quad .
\end{equation}

All this indicates that correlation effects are strongly enhanced at
the surface relative to the inner layers, in good agreement with
experimental observations for CaVO$_3$ and the closely related
material SrVO$_3$
\cite{Maiti_et_al:2001,Eguchi_et_al:2006,Laverock_et_al:2013,Laverock_et_al:2015}
and previous DFT+DMFT calculations for SrVO$_3$
\cite{Liebsch:2003,Ishida/Wortmann/Liebsch:2006}.
In particular, close to \umit\ but still in the metallic phase, the
surface seems to form a ``dead layer'', with nearly vanishing
quasiparticle weight, as discussed in
Ref.~\onlinecite{Borghi/Fabrizio/Tosatti:2009} based on DMFT
calculations for a one-band Hubbard model within a slab geometry.

Finally, we also consider the combined effect of finite thickness and
tensile epitaxial strain. We focus on the case of a 3-bilayer slab
with 4\,\% tensile strain, shown in \pref{fig:DMFT46l}c.
Similar to the case of the strained bulk systems, tensile strain leads
to a strong shift of \umit\ to lower values also in the strained
3-bilayer slab, resulting in an insulating state at $U=5\ev$.
Moreover, comparing to the bulk reference at 4\,\% tensile strain, one
can see that the effect of the reduced thickness is rather weak, as
was the case in the corresponding unstrained system.
Again, the surface layer exhibits a more continuous change of
$\bar{A}(0)$ but maintains essentially the same \umit\ as the inner
layers.
From the orbital occupations shown in \pref{fig:OCC6l}, it can be seen
that in the strained slab all layers are strongly polarized in the
insulating state. Thereby, the behavior of the inner layers is very
similar to the corresponding strained bulk system (compare
\pref{fig:OCC6l}b with \pref{fig:bulkDMFT}b) whereas the surface layer
is already strongly polarized in the metallic phase.
This is again fully consistent with the calculated crystal-field
splitting between the MLWFs, shown in \pref{fig:CFS}b. The splitting
in the inner and subsurface layers of the strained 3-bilayer slab is
very similar to the bulk splitting under 4\,\% tensile strain, while
in the surface layer this splitting is further enhanced.

It thus appears that in all cases the MIT is mainly controlled by the
local crystal-field splitting within the \ttg\ manifold and the
resulting tendency to form an orbitally polarized insulating state. As
seen in \pref{fig:CFS} the crystal-field splitting in all layers is
strongly affected by tensile epitaxial strain, whereas the finite size
effects seem to manifest mostly as an enhanced crystal-field splitting
in the surface layer. The insulating character of the unstrained
1-bilayer slab therefore appears more like a surface-induced effect
rather than being related to the formation of quantum well states
along the $c$ direction. Even though a peak structure related to such
quantum well states is clearly visible in the DFT-DOS of the inner
layers of the 3-bilayer slab (see \pref{fig:pdos}), this does not seem
to have a big impact on the metallic character.

We note that in Ref.~\onlinecite{Sclauzero/Dymkowski/Ederer:2016} we
found that, while the crystal-field splitting can nicely explain the
trends in \umit\ for a $d^1$ system under tensile strain, the
strain-induced changes in the hopping amplitudes also have to be taken
into account to obtain the correct magnitude of the shift in \umit. In
contrast, for the finite size effect discussed here, the hopping
amplitudes appear to be less relevant, since, as noted previously,
they do not exhibit any significant changes with reduced thickness.

\section{Summary and conclusions}

In summary, we have used DFT+DMFT calculations to compare the effects
of tensile epitaxial strain and finite film thickness on the tendency
of \cvo\ to undergo a Mott MIT.
We find that tensile strain strongly decreases \umit, the critical
interaction strength for the Mott transition, and that strains of the
order of what is achievable in a fully strained \cvo\ film grown on
\sto\ (about 3.5\,\%) are sufficient to induce a transiton to the
Mott-insulating state.
Alternatively. a transition to the insulating state can also be
achieved in ultrathin free-standing slabs of \cvo, however, only for
thicknesses smaller than approximately 4 perovskite units. In
contrast, a 6-layer slab already exhibits bulk-like behavior.

Furthermore, we observe a strong suppression of the quasiparticle
weight in the surface layer, consistent with previous experimental and
theoretical reports for CaVO$_3$ and SrVO$_3$. This is due to a strong
surface-induced crystal-field splitting, which results in strong
orbital polarization inside the surface layer. Our results show that
these modifications are essentially restricted to the outermost layer
and that the bulk properties are nearly recovered already in the
subsurface layer.

Our results demonstrate that the effects of tensile epitaxial strain
and the finite size or surface effects cooperate, but that the strain
effect clearly dominates, except in the ultra-thin limit (i.e., below
4 perovskite layers). Since in Ref.~\onlinecite{Gu_et_al:2013}
insulating behavior was observed for a thickness of up to 4\,nm, i.e.,
10-11 perovskite layers, this suggests that factors other than pure
finite size effects are at play here. The scenario outlined in
\pref{sec:intro}, i.e., a strain-induced insulating state in the fully
strained thinnest films, which evolves back to metallic bulk behavior
in thicker films where the epitaxial strain is relaxed, provides a
potential alternative explanation for these experimental observations.
However, other factors, not considered in this work, could also be of
importance, such as, e.g., point defects, interface roughness, or
other forms of disorder, which could then also lead to an insulating
state via Anderson localization. Further experimental studies are
required to clarify this.
Nevertheless, we point out that the thickness of approximately 4
perovskite units, for which we find the \cvo\ slabs to be just at the
border of the MIT, is in very good agreement with the thickness of 2-3
monolayers, for which a gap opening was observed by photoemission
spectroscopy in SrVO$_3$ films grown by molecular beam epitaxy
\cite{Yoshimatsu_et_al:2010}.
Furthermore, in Ref.~\onlinecite{Liao_et_al:2015} it was shown that
the critical thickness for the formation of an insulating state in
La$_{2/3}$Ca$_{1/3}$MnO$_3$ can be reduced to only 3 unit cells by
minimizing the substrate-induced strain as well as the amount of
oxygen vacancies.
Thus, our results for the unstrained slabs can be considered as
providing a lower limit for the thickness above which a metallic state
can be preserved in \cvo, as long as other effects promoting the
insulating state can be excluded.

Finally, our study shows that it is possible to tune the Mott
transition in \cvo\ (and related materials) through multiple control
parameters. This provides an exciting avenue for the design of Mott
transistors \cite{Newns_et_al:1998,Inoue/Rozenberg:2008} and other
``Mott-tronic'' devices \cite{Yang/Ko/Ramanathan:2011}, or also to
tune the properties of correlated metals for other potential
applications, e.g., as transparent conductors \cite{Zhang_et_al:2016}.

\acknowledgments

This research was supported by the NCCR MARVEL, funded by the Swiss
National Science Foundation. The authors thank Daniel McNally, Milan
Radovic, and Thorsten Schmitt for insightful discussions. Calculations
have been performed on ``Piz Daint'' at the Swiss National
Supercomputing Centre.

\bibliography{references}

\end{document}